\documentclass[aps,pre,onecolumn,12pt]{revtex4-1}
\usepackage{amsmath}
\usepackage{amssymb}
\usepackage{graphicx}
\usepackage{color}
\begin{document}
\title{Statistics of co-occurring keywords on Twitter}
\author{Joachim Mathiesen$^1$, Luiza Angheluta$^2$ and Mogens H. Jensen$^1$}
\affiliation{$^1$Niels Bohr Institute, University of Copenhagen, Blegdamsvej 17, DK-2100 Copenhagen, Denmark\\
$^2$Physics of Geological Processes, Department of Physics, University of Oslo, P.O. 1048 Blindern, 0316 Oslo
Norway}
\begin{abstract}
Online social media such as the micro-blogging site Twitter has become a rich source of real-time data on online human behaviors. Here we analyze the occurrence and co-occurrence frequency of keywords in user posts on Twitter. From the occurrence rate of major international brand names, we provide examples on predictions of brand-user behaviors. From the co-occurrence rates, we further analyze the user-perceived relationships between international brand names and construct the corresponding relationship networks. In general the user activity on Twitter is highly intermittent and we show that the occurrence rate of brand names forms a highly correlated time signal. 
\end{abstract}
\maketitle

\section*{Introduction}
Human activity is often intermittent with periods of low activity punctuated by bursts of high activity \citep{B05,VOD06}. Events in society often animate people on larger scales reflecting a tendency for alignment (i.e.~correlation) between individuals. These correlations can be observed in different social organizations and directly measured in online social media such as Twitter or in stock market data. It is well known that breaking news can make stock traders collectively rush in or out of certain stocks. Similarly, events in society can generate a surge of new user posts on Twitter. In both cases, the activity levels have recently been shown to share common statistical properties \citep{MAAJ13}. Intermittent human dynamics have been observed in individuals browsing internet sites\citep{GR08}, in the online blogosphere \citep{LKG07} and during crowd panic \citep{HFV00,HJM06, DJS06}. Correlation in behaviors has also been shown to be a consequence of time resource management\citep{HJM12}.

Twitter is a microblogging service where registered users can post short text messages up to 140 characters in length. These text messages are known as ``tweets''.  The frequency by which users on a global scale post information on social media allows for a high-resolution real-time analysis of user behavior. The Twitter user base is predominately located in the USA~\citep{KKN12} and might not be a perfect representation of the general public~\citep{MLA11}.

Here we shall consider the statistics of keyword occurrence and co-occurrence on Twitter. From this statistics it might be possible to infer basic properties of human behavior in online communities. Observations on collective human behavior has been used in predictions of future developments in society~\cite{CV12} and and the public sentiment, as extracted from tweets, has been used in prediction of movements on the stock market~\citep{B11}. Specifically, the variations in the score of public happiness and calmness were shown to be predictive for movements of the Dow Jones Industrial Average index. A more basic correlation was found between the occurrence frequency of certain words on Twitter and the up and down turns of stock prices \citep{ZFG11}. The sheer number of tweets mentioning specific movies has also been shown to be predictive for box office sales \citep{AH10}.

One way to analyze the user activity on Twitter is to measure the frequency by which certain words are mentioned. Examples of words with a global appeal could be international brands names like Apple, IBM, Starbucks, etc. For each of these brand names, the occurrence rate in tweets is characterised by days of steady levels, which are interrupted by spikes of high activity.  When a company launches a new product, a number of people will immediately be tweeting about it, and thereby encourage others to share their experience with the product, which quickly creates a global awareness. In fact, an analysis of activity levels shows that this awareness, quantified by increased activity levels on Twitter, has a typical persistence of about 24 h after which it slowly fades away. This persistence or memory effect is observed both on social media and in financial markets \citep{MAAJ13}.

\section{Results}

\subsection{Dynamics of tweet rates}

The occurrence rate of new tweets containing a specific word is estimated as follows. A tweet, $T_i$, returned from a query to Twitter contains a list of information $T_i=(s,t_i,\ldots)$ including the tweet text string $s_i$, the time $t_i$ when the tweet was posted, a user id of the user and further information that is not used in the present work.

A time-signal, $\eta_a(t)$, is then formed from Tweets mentioning a specific word $a$ by the following sum
\begin{equation}\label{pp}
\eta_a(t)=\sum_i \delta(t-t_i).
\end{equation}
From the signal, we can then compute an averaged tweet rate from an unnormalized kernel density smoother,
\begin{equation}
\gamma_a(t)=\frac 1 {\tau} \int_{-\infty}^{\infty}K\left(\frac{t-t'}\tau\right)\eta_a(t') \mathrm{d} t',  \label{tweetrate}
\end{equation}
In the following, we shall use a rectangular integration kernel $K(x)$ of unit height and width such that the rate simply is equal to the number of tweets in a time window $\tau$, ${n_a(t)}/{\tau}$. Similarly we might define a rate for tweets containing multiple words of interest, say how frequently do $a$ and $b$ co-occur in a tweet,
\begin{equation}\label{dualrate}
\gamma_{ab}(t)={n_{ab}(t)}/{\tau}.
\end{equation}
The data used in the current analysis has in part been used in Ref.~\citep{MAAJ13}.

\begin{figure}
\includegraphics[width=.5\textwidth]{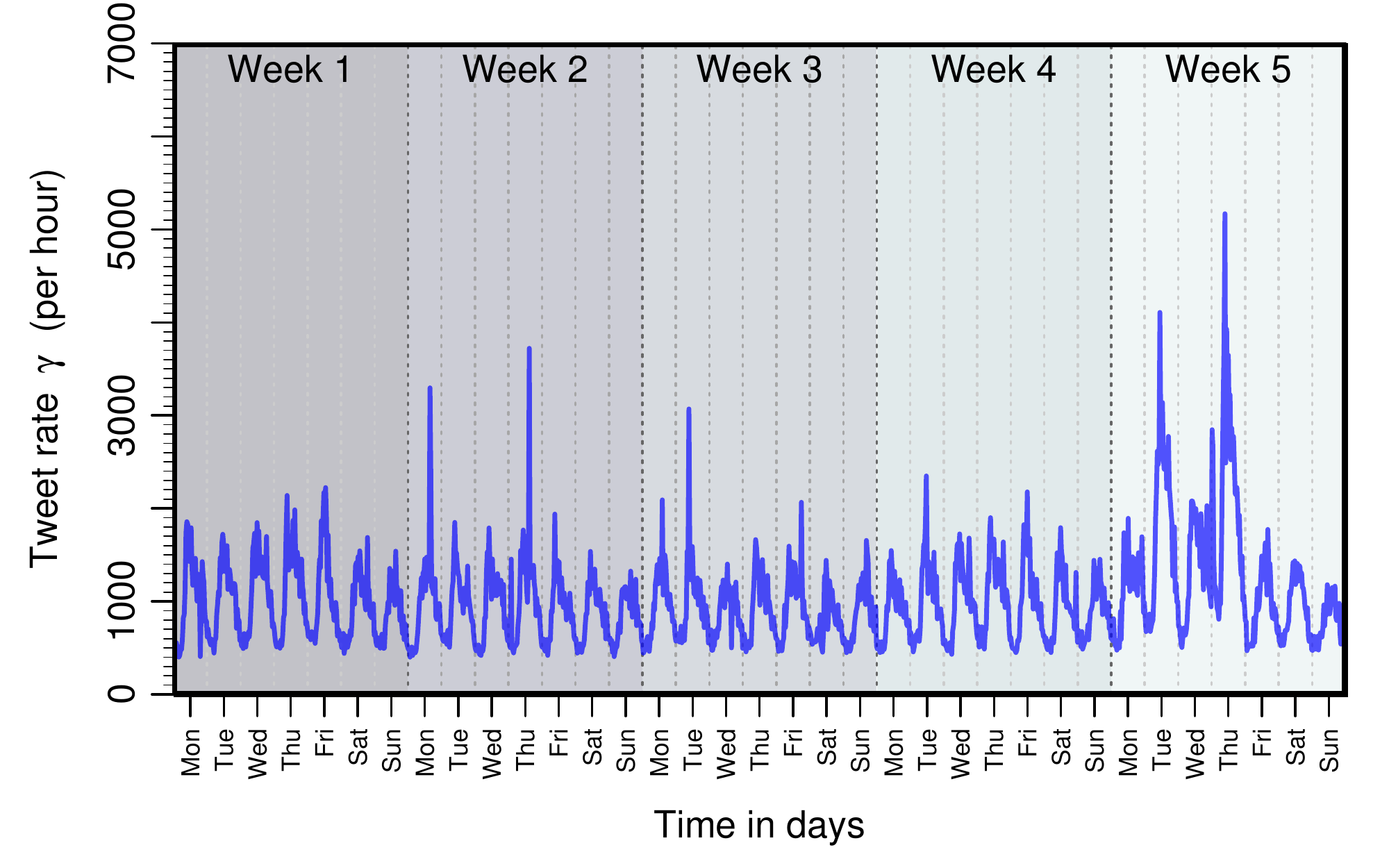}
\caption{Time signal formed from tweets mentioning "Starbucks". The signal consists of an underlying regular signal punctuated by bursts of larger user activity. Time is shown in Eastern Standard Time (EST). Note that the rate is limited by the fraction of new Tweets made available by Twitter.}\label{starbucks}
\end{figure}

In Fig.~\ref{starbucks} we show the variation in the rate of tweets mentioning "Starbucks" over a three week period. The time signal has a clear 24 hour cycle as well as a weekly variation. On top of these variations, short bursts of larger activities can be seen. The statistics of the magnitude of these bursts follows a broad distribution as we shall discuss below.

\begin{figure}
\includegraphics[width=.5\textwidth]{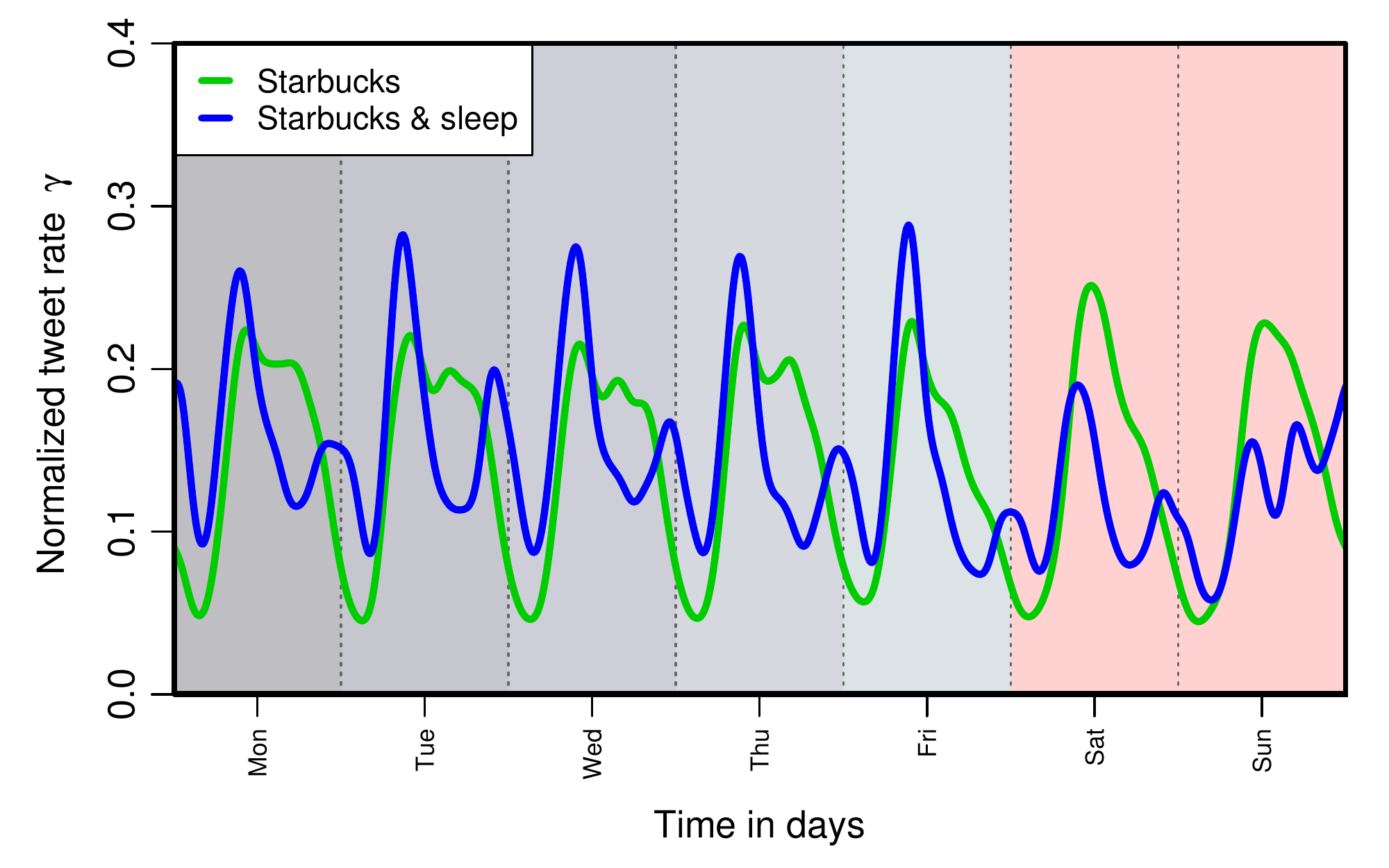}
\caption{Averaged and normalized time signals for tweets mentioning "Starbucks" (green line) and for tweets mentioning simultaneously "Starbucks" and "sleep" (blue line). We see that the latter time-signal has a characteristic double peak during the weekdays where people mostly mention sleep in the morning and in the night. The fraction of "Starbucks" tweets mentioning also "sleep" is around 0.5\%. Time is shown in EST.}\label{sleep}
\end{figure}

\begin{figure}
\includegraphics[width=.5\textwidth]{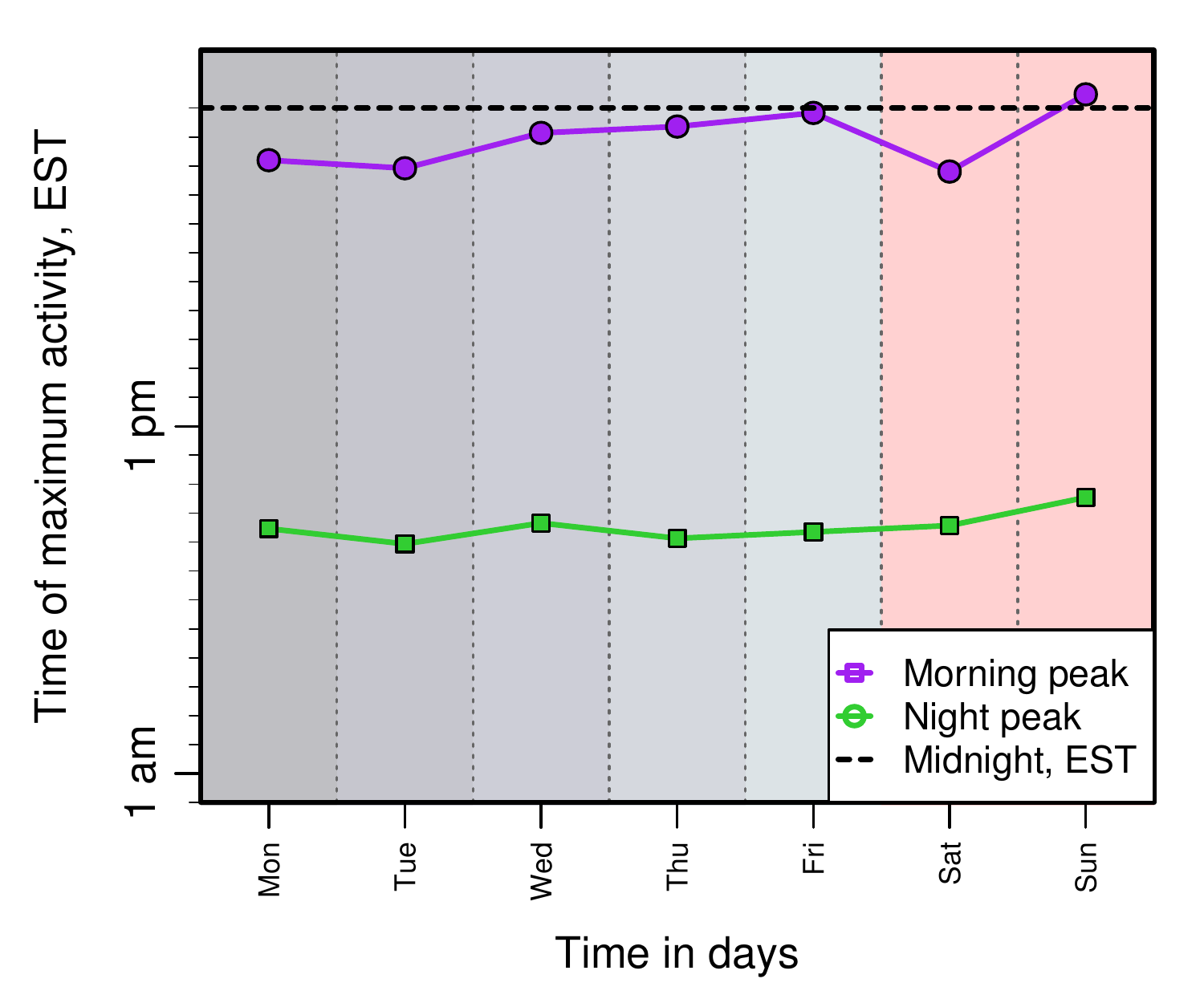}
\caption{Time of the day where the maximum rate is reached of tweets mentioning simultaneously "Starbucks" and "sleep". The morning peak is marked with "squares" and the evening peak with "circles". Time is again shown in Easter Standard Time. We see that time of maximum activity  slightly increases towards the weekends.}\label{sleeptime}
\end{figure}

The Twitter time signals provide basic insight on human habits. In fig.~\ref{sleep} we show the co-occurrence rate of tweets mentioning simultaneously "Starbucks" and "sleep" (including related words like asleep, sleeping, sleepy ...). Relative to the signal of tweets mentioning only "Starbucks" we see a pronounced peak both in the morning and in the night. While tweets in the morning often mention the need or joy of coffee the peak in the evening is in part due to people tweeting about problems falling asleep after drinking coffee. In general sleep seems to be less a problem in the weekends where the frequency of tweets mentioning sleep drops while the overall time signal for "Starbucks" remains unchanged. We also see in Fig.~\ref{sleeptime} that the time at night where sleep is mentioned most frequently slightly increases throughout the week.

\begin{figure}
\includegraphics[width=.5\textwidth]{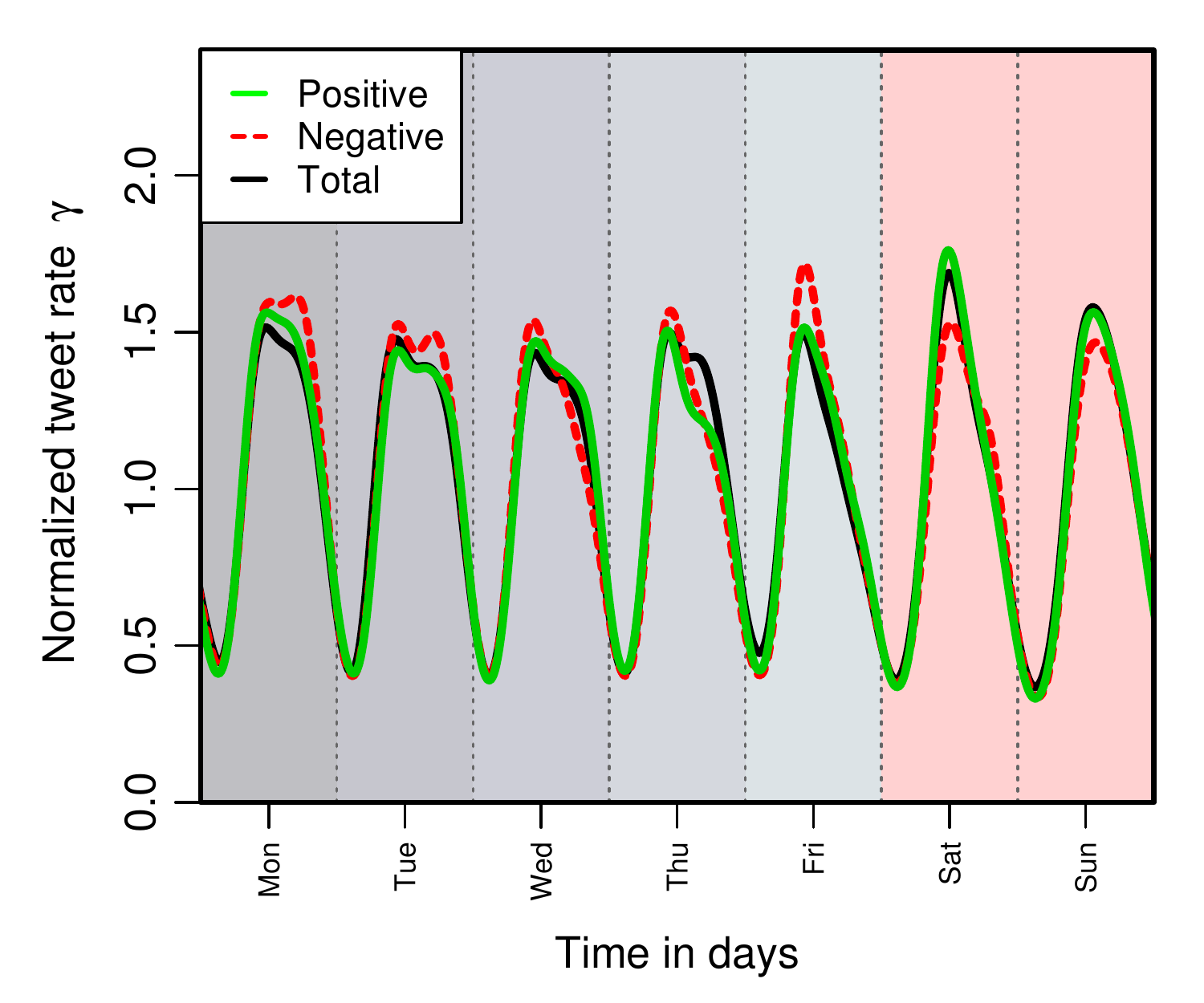}
\caption{Normalized rates of tweets with positive ($s>0$) and negative ($s<0$) sentiment scores. There is relatively more tweets with a positive score in the weekend, whereas the weekdays are relatively more negative. Note that the average sentiment score of all tweets is approximately $0.2$ and that there is approximately twice as many tweets with a positive score than with a negative score. The signals have been smoothed with a bandwidth of 2hrs. }\label{sentiment}
\end{figure}

Moreover we may perform a sentiment analysis of the user posts. From a basic text analysis of all tweets mentioning a specific brand, we count the number of words with a negative $n_-$ and positive $n_+$ sentiment according to a pre-compiled list of words. The difference, $s=n_+-n_-$ between the number of positive and negative words is then used to form a sentiment score $s$ of the individual tweets. We show in Fig.~\ref{sentiment} the normalized tweet rate $\gamma/\bar{\gamma}$ of tweets (where $\bar{\gamma}$ is the average tweet rate) with a positive and negative sentiment score, respectively. There is a relatively higher density of tweets with a negative sentiment in weekdays whereas the mood changes in the weekend and the positive sentiment has a higher density.

\begin{figure}
\includegraphics[width=.5\textwidth]{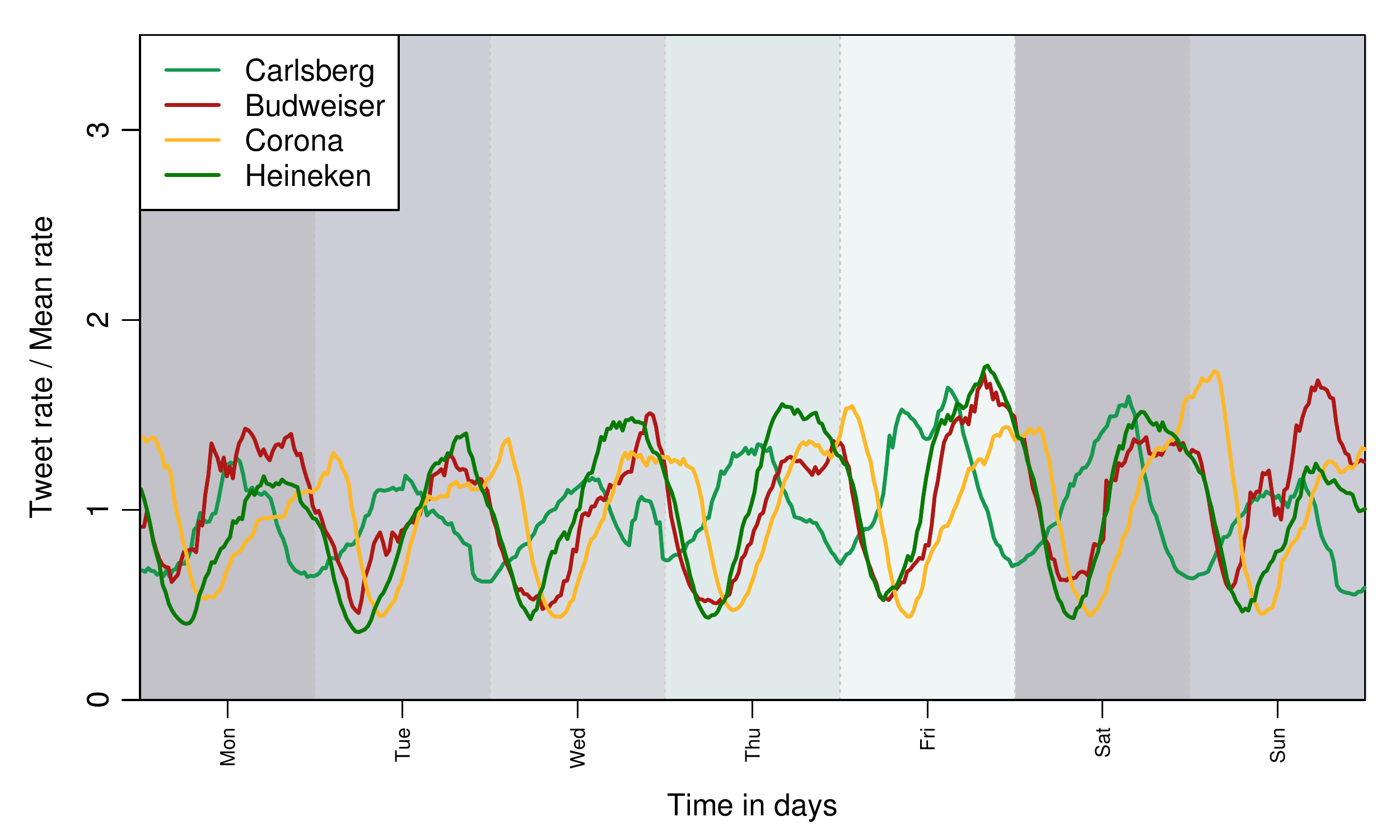}
\caption{Normalized tweet rates for four different beer brands Budweiser, Carlsberg, Corona and Heineken. A Clear phase shift is observed in the time signals reflecting a difference in the primary markets of the beers, i.e Budweiser and Corona are correlated with American time zones while Carlsberg is correlated with Asian time zones.}\label{beers}
\end{figure}

The Twitter user base is to a large extent located in the USA and the activity pertaining to brands like Starbucks correlate strongest with American time zones. However, certain brands have larger markets outside the USA and do therefore follow more closely day and night cycles of other time zones. In Fig.~\ref{beers} we show a normalized tweet rate for four different beer brands, Budweiser, Carlsberg, Corona and Heineken. We see that there is clear shift in the time of maximum tweet rate for the brands reflecting a difference in primary markets of the brands. We see that Carlsberg is shifted more towards European and Asian time zones relative to e.g. Corona which correlate strongly with American time zones.

\section{Network of brands}
Based on the co-occurrence rate of how often two different brands occur in the same tweet, we now analyse further how individual brands are linked to each other. With this information one can construct a network of important international brands.  Eq.~(\ref{dualrate}) can be used to define a normalized symmetric measure of similarity (the Jaccard index) by dividing the co-occurrence rate with the combined rate of seeing at least one of the brands in a tweet,
\begin{equation}
\omega_{ab}=\frac{\gamma_{ab}}{\gamma_{a}+\gamma_{b}-\gamma_{ab}}\label{measure}
\end{equation}

\begin{figure}
\includegraphics[width=.7\textwidth]{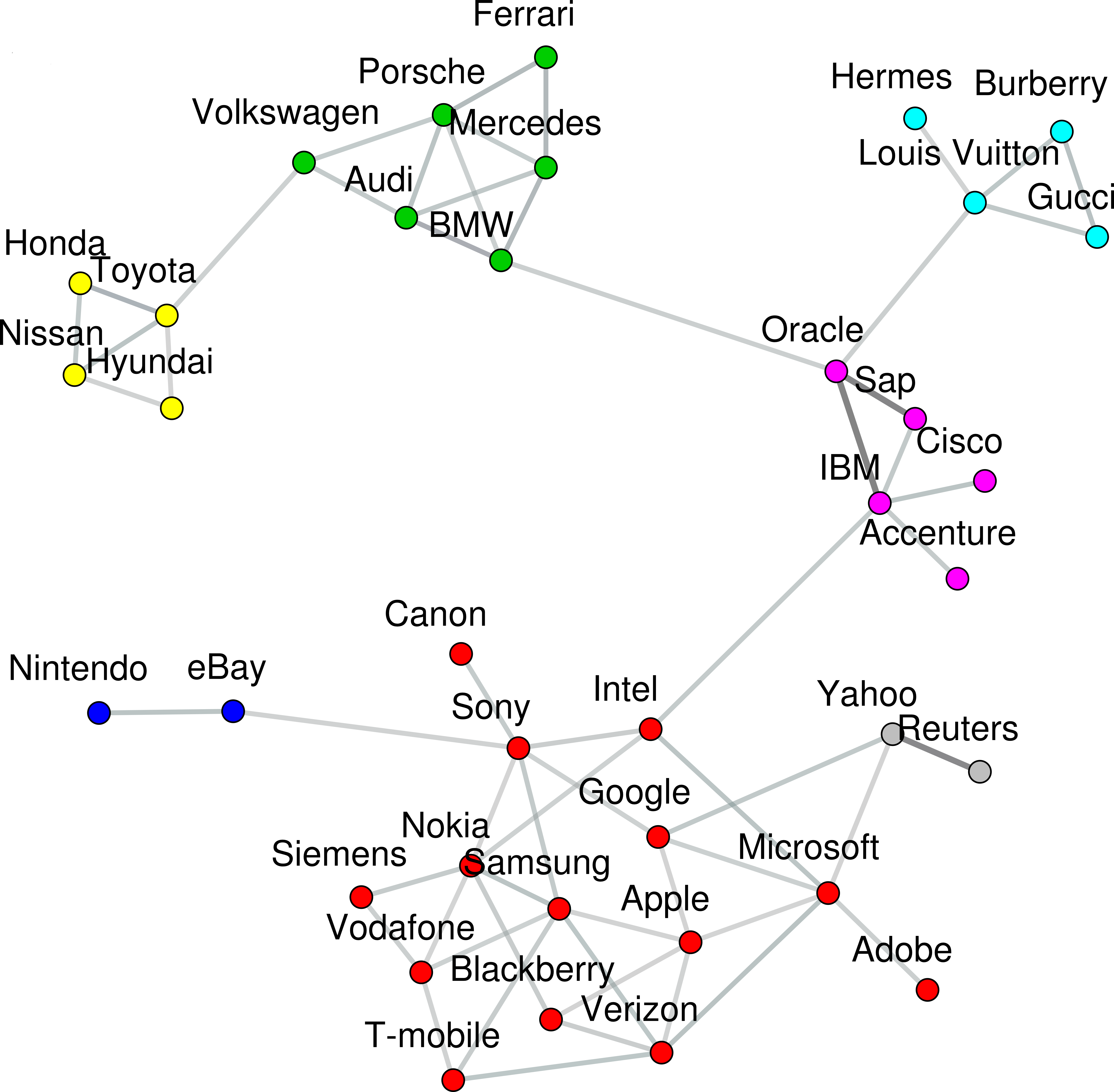}
\caption{Figure taken from Ref.~\cite{MYJ12}. Network of international brands names. The links strengths are calculated from Eq.~(\ref{measure}) and are based on the co-occurrence rate brand names in user posts on Twitter. We use a threshold value $\omega_{AB}<0.004$. }
\label{network}
\end{figure}

In Fig.~\ref{network} we present the network of the international brands where the link strength is given by the measure Eq.~(\ref{measure}). A threshold is introduced on the link strength in order to visualize primary structures, {\it i.e.}~links between brand with a similarity $\omega_{AB}<0.004$ are omitted. We observe that the network is strongly modular with groups of brands representing similar products or services. As an example one can observe distinct groups of European car brands, Asian car brands, consulting and IT companies, and fashion brands. The modules in the network are coloured according to the community detection algorithm introduced in~\cite{rosvall}. Most of the connections inside the modules are rather obvious, whereas a few links connecting the modules represent less obvious relations between brands.

\begin{figure}
\includegraphics[width=.5\textwidth]{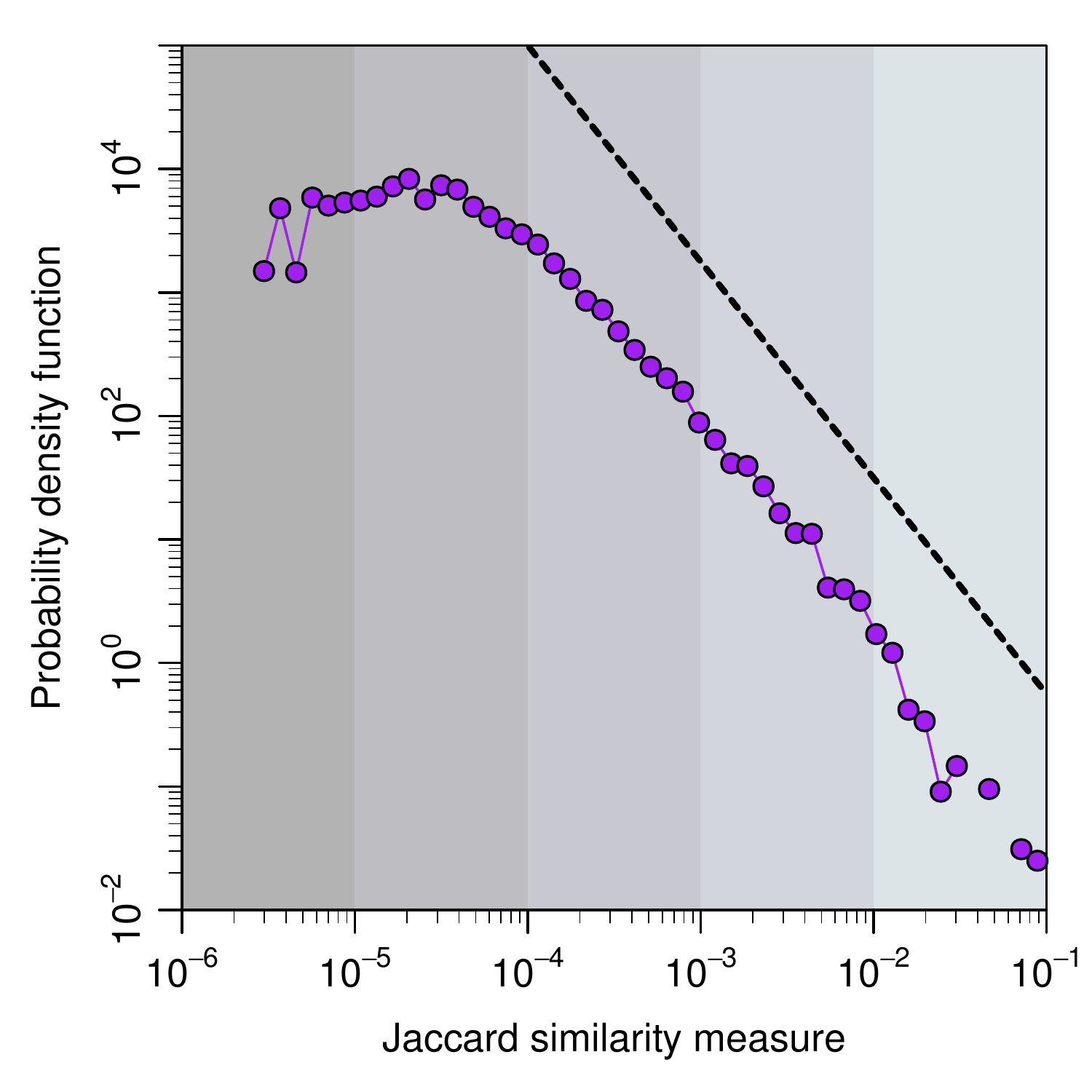}
\caption{The distribution of the link strength between international brands names computed from Eq.~(\ref{measure}). The dashed line is a guide to to the eye with and corresponds to a scaling exponent of $\alpha=-1.7$.}
\label{ls}
\end{figure}

We now consider the distribution, $p(\omega_{ab})$, of the link strengths which are shown in Fig.~\ref{ls} on double logarithmic axes and appears to follow a power law,
\begin{equation}
p(\omega_{ab})\sim\omega_{ab}^{-\alpha},
\end{equation} 
with a scaling exponent $\alpha=1.7$. In general, the tweet rate of pairs $\gamma_{ab}$ does not follow trivially from the rate of the individual brands~\citep{MYJ12}. In fact, the modular structure where a few brands are strongly connected is contributing strongly to the broad tail of the distribution. A power-law distribution has also been observed in related semantic network \citep{ST05} and for the co-occurrence of tags in social annotation systems \cite{CB09} where users annotate online resources such as web pages by lists of words. However, the scaling exponent in the latter case ($\alpha>2$) is larger than the one reported here.

\section{Statistics of the activity levels}
The user activity levels on Twitter have been suggested to be driven predominantly by external events rather than direct user interactions \citep{MAAJ13}. In other words, we may in a simple model of activity levels neglect the network formed between the Twitter users. We shall model the tweet dynamics by a stochastic point process, Eq.~(\ref{pp}), where tweets occur at random times separated by random time intervals. If the time interval between successive tweets fluctuates around a constant mean value, then the point process corresponds to a Poisson process and the time interval would be exponentially decaying away from the mean. However, non-trivial dynamics and power-law statistics can be obtained when the average waiting time, $\tau$, is itself a stochastic variable. In that way, the statistics of all quantities of interest depends entirely on the underlying stochastic dynamics of the average waiting time.

Without correlations between user activities, the tweet rate $\gamma(t)=1/\tau$ is determined by a balance between users becoming inactive and new users activated by external events. We shall assume that the correlations enter in the shape of a multiplicative noise with an amplitude given by $\sqrt\tau$, 
\begin{equation}\label{eq:ABBM}
\frac{d\tau}{dt} = -f(\tau)+c+\sqrt{\tau}\xi(t), 
\end{equation}
where $f(\tau)$ is an external drive that keeps users active and thereby the waiting time low, $c$ is the rate at which users become inactive, and $\xi(t)$ is a fluctuating noise with zero mean and unit variance 
\begin{equation}
\langle \xi(t)\xi(t')\rangle = \delta(t-t').
\end{equation}
Eq.~(\ref{eq:ABBM}) is here interpreted in the Ito sense. Below we shall assume a simple form for the driving force $f(\tau)=k\tau$, where $k$ is a constant. The Fokker-Plank equation associated with Eq.~(\ref{eq:ABBM}) then reads
\begin{equation}\label{eq:FP_ABBM}
\frac{\partial}{\partial t}P(\tau,t)= -\frac{\partial}{\partial \tau}\left((-k \tau+c\right)P(\tau,t))+\frac{1}{2}\frac{\partial^2}{\partial \tau^2}\left(\tau P(\tau,t)\right), 
\end{equation}
with the steady state probability distribution function (PDF)
\begin{equation}
P(\tau)\sim \tau^{-1+2c}\exp(-2k \tau).
\end{equation}
We notice that for $c=0$, the PDF has a non-integrable singularity at $\tau=0$. Therefore $P(\tau)$ is not normalizable and the only stationary solution is $P(\tau) = \delta(\tau)$. For $c>0$, Eq.~(\ref{eq:ABBM}) has a repelling wall at $\tau=0$ and the PDF attains a stationary solution that can be normalized. 

In general, the average waiting time $\tau$  between tweets is assumed to increase with time $\partial_t \tau=1$ when there is no external driving force. We then achieve the average waiting $\tau$ time between tweets by setting $c=1$, i.e. 
\begin{equation}
P(\tau)\sim \tau \exp(-2k \tau),
\end{equation}
thus there is a small probability both for very short and very long waiting times with a maximum likelihood at a finite, intermediate value $\langle \tau\rangle = 1/k$. Notice that the average waiting time diverges as $k\rightarrow 0$. The important thing is that the PDF is linearly proportional to small $\tau$'s. This implies that the tweet rates which is $\gamma \sim 1/\tau$ is described by a PDF 
\begin{equation}\label{scaling}
P_\gamma(\gamma) = \left|\frac{d\tau}{d\gamma}\right| P(\tau) \sim \gamma^{-3} \exp(-2k/\gamma).
\end{equation}
Hence, the PDF $P(\tau)\sim \tau$ at small $\tau$'s corresponds to $P_\gamma(\gamma)\sim \gamma^{-3}$ at large $\gamma$'s. 
 
\begin{figure}
\includegraphics[width=.5\textwidth]{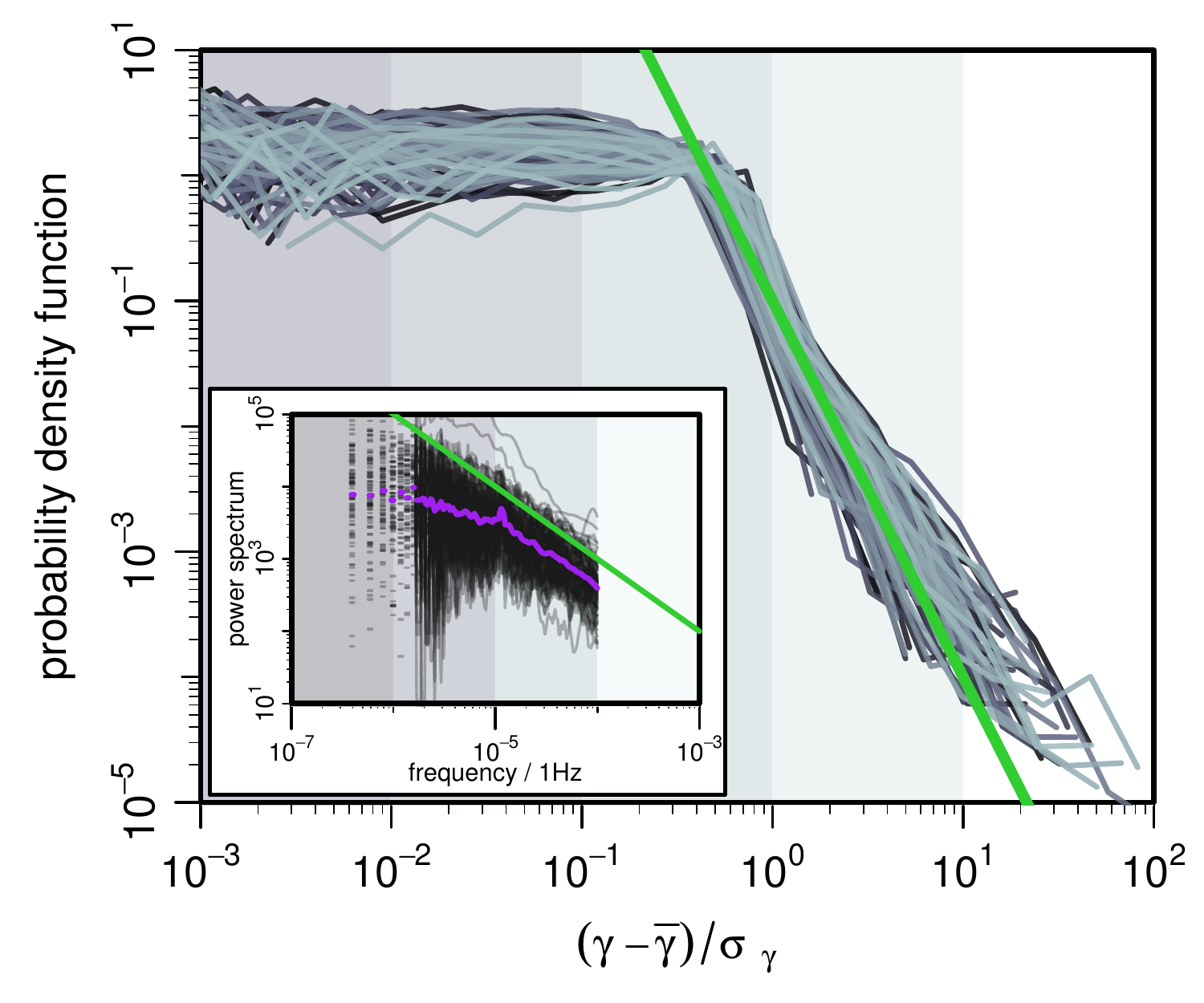}
\caption{The distribution of Tweet rates for 92 international brand names. The distributions follow a power law (represented by the line added as guided to the eye) in the tail with a scaling exponent $\alpha\approx-3$. In the inset we show the corresponding power spectra which scales inversely with the frequency, i.e.~as $1/f$. Again the line correponds to $1/$ scaling.}\label{1f}
\end{figure}

Interestingly, these two signals have very different power spectrum densities: $S_\tau(f)\sim f^{-2}$, while $S_\gamma(f)\sim f^{-1}$. In Fig.~\ref{1f} the distributions of the occurrence rates of tweets mentioning the major brand names (see \citep{MAAJ13} for the complete list of brands used). We observe that the occurrence rates follows the same power law in the tail with an exponent close to the one predicted from our model in Eq.~(\ref{scaling}). Similarly do the corresponding power spectra reveal a characteristic $1/f$ scaling.

\subsection{Numerical scheme} 
These predictions can also be checked by a direct numerical integration of Eq.~(\ref{eq:ABBM}). In a regular time discretization, the noise term would introduce negative values, which are forbidden in the continuum. An efficient and accurate numerical method is based on a split-operator scheme where the square Bessel-process is separated from the rest of the dynamics. The square-Bessel process models the Brownian motion in $d $-dimensional space and has the $\sqrt{x}$-multiplicative noise. The transition probability for this process is exactly known and given by the non-central $\chi^2$-distribution (the central $\chi^2$ distribution is a special case of the $\Gamma$ distribution). 

\section{Concluding remarks}
In this paper we have used data from the social media Twitter in order to analyse the occurrence and co-occurrence rate of keywords. The analysis provides information about basic behavior of Twitter users and to some extent brand-users. Furthermore, our analysis might be of commercial interest, in the sense that repeated brand advertisement campaigns probably would have a larger impact at the time of week when there is a maximum awareness in the public about the brand. As we observed, the user awareness varies throughout the day and week. The co-occurrence rates of brands might also provide important information about competitors in the market. Finally, we have considered the fluctuations in the combined activity levels of the social media users and we believe that these fluctuations tell something basic about human behavior. Without question social media provides an excellent source for further studies along these lines.

\section*{Acknowledgments}
Sune Lehmann is gratefully acknowledged for useful comments and suggestions. This study was supported by the Danish National Research Foundation through {\it the Center for Models of Life} and by the UCPH 2016 (University of Copenhagen) funds through the {\it Social Fabric} grant.

\end{document}